\documentclass[lettersize, journal, 10pt]{IEEEtran}
\IEEEoverridecommandlockouts
\usepackage{graphicx}
\usepackage{subcaption}
\usepackage{cite}
\usepackage{amsmath,amssymb,amsfonts}
\usepackage{algorithmic}
\usepackage{algorithm}
\usepackage{array}

\usepackage{textcomp}
\usepackage{xcolor}
\usepackage{url}
\usepackage{verbatim}
\usepackage{cite}
\usepackage{stfloats}

\usepackage{multirow}
\usepackage{multicol}
\usepackage{comment}

\begin{document}

\title{Design of a Superconducting Multiflux Non-Destructive Readout Memory Unit}

\author{\IEEEauthorblockN{Beyza Zeynep Ucpinar, Yasemin Kopur, Mustafa Altay Karamuftuoglu, Sasan Razmkhah, and Massoud Pedram}\\
\IEEEauthorblockA{\textit{Ming Hsieh Department of Electrical and Computer Engineering} \\
\textit{University of Southern California}\\
Los Angeles, USA \\
\{ucpinar, kopur, karamuft, razmkhah, pedram\}@usc.edu}
}

\maketitle
\begin{abstract}
Due to low power consumption and high-speed performance, superconductor circuit technology has emerged as an attractive and compelling post-CMOS technology candidate. However, the design of dense memory circuits presents a significant challenge, especially for tasks that demand substantial memory resources. While superconductor memory cells offer impressive speed, their limited density is the primary yet-to-be-solved challenge.
This study tackles this challenge head-on by introducing a novel design for a Non-Destructive Readout (NDRO) memory unit with single or multi-fluxon storage capabilities within the same circuit architecture. Notably, single storage demonstrates a critical margin exceeding 20\%, and multi-fluxon storage demonstrates 64\%, ensuring reliable and robust operation even in the face of process variations. These memory units exhibit high clock frequencies of 10GHz. The proposed circuits offer compelling characteristics, including rapid data propagation and minimal data refreshment requirements, while effectively addressing the density concerns associated with superconductor memory, doubling the memory capacity while maintaining the high throughput speed.

\end{abstract}
\begin{IEEEkeywords}
Superconductive electronics, Rapid single flux quantum memory, multiple-bit storage cell, non-destructive read-out memory
\end{IEEEkeywords}

\section{Introduction}
Superconducting circuit technology offers distinct advantages over traditional semiconductor-based circuits, particularly in terms of power efficiency and ultra-high speed \cite{likharevRSFQ}. In this context, it holds significant promise as a future complementary technology to CMOS, considering that Moore's Law is no longer applied to semiconductor technology. However, developing dense memory remains a primary challenge with superconductor technology. Superconductor logic circuits invariably require storage units to facilitate data propagation and processing. Even though superconductor memories are low power consuming and have high-speed operation, they also have disadvantages with complex cooling systems, underdeveloped fabrication processes, and low implementation density. Previous research has explored the use of shift registers to achieve 1-kbit and 2-kbit memory capacities \cite{mukhanov_shift_reg_93, yuh_2kb_shift_reg_93}. While the shift register-based superconductor memory architecture exhibits commendable characteristics, the inherent write-back cycle overhead within the design gets expensive with scale.

Various approaches to Random Access Memory (RAM) designs have been proposed. They include those utilizing DC-powered designs \cite{kirichenko_ram_dc_powered}, shift register based \cite{tanaka_ram2016}, and hybrid implementations \cite{hybrid_sup_mag}. Implementations including Magnetic Josephson Junctions (MJJs) \cite{mukhanov_ferromagnetic, mjj_mem, magnetic_jj_croyogenic} have plenty of advantages with fast switching, promising high density and non-volatile features. However, developing materials and related fabrication remains challenging. Furthermore, kinetic inductance-based methodologies, as demonstrated by studies \cite{kinetic_inductance_0, Murphy_2017_kinetic_inductance, kientic_inductance_2}, have yielded several encouraging outcomes. At the same time, challenges related to integration, materials selection, and fabrication processes remain pertinent issues within this research domain. These efforts aim to identify potential candidate implementations for cryogenic memory devices \cite{cryogenic_mem_tech}. 

The fundamental element of dense memories in all studies is the memory unit's design, whether for one-bit or multiple-bit storage. Multiple-bit storage unit enables high-density storage. Efficient multi-flux storage has been demonstrated in High Capacity Destructive Readout (HC-DRO) \cite{ndro_haipeng} and scalable 1-bit Non-Destructive Readout (NDRO) cell designs \cite{Katam_2020_ndro}.

This study introduces two memory unit designs: a 1-bit NDRO unit and a 2-bit Multi-NDRO (M-NDRO) unit, both featuring a local feedback wiring structure. The proposed architecture draws inspiration from Dynamic Random Access Memory (DRAM) designs, implementing a local feedback wiring circuit that enables immediate data reloading after a destruction readout. Leveraging the high-speed propagation of Single Flux Quantum (SFQ) pulses in superconducting circuits, we effectively minimize data refreshment times to negligible levels.

In the single-bit NDRO circuit, a single flux represents 1 bit of data. Conversely, the M-NDRO circuit utilizes three fluxons to represent 2-bit data. The overall layout size remains consistent despite differences in their input/output (I/O) configurations and memory unit structures. The M-NDRO circuit incorporates a Multiple-Clock Generator (MCG) circuit to facilitate the representation of 2-bit data. Additionally, depending on the presence of the Reset Generator (RG) unit, the M-NDRO cell supports either the resetting of the storage or the decrementation of the stored fluxes by one.

Both memory unit designs share identical structures, with parameter value variations determining their flux storage characteristics. The storage properties of a superconductor loop are determined by its current and inductance. Although the M-NDRO circuit requires additional units, namely the multiple-clock generator and reset generator units, the local feedback wiring structures remain consistent for both designs. These feedback structures consist of fundamental wiring cells, including Josephson Transmission Line (JTL), Splitter (SPL), and Confluence Buffer (CBU) units, all of which we have optimized for minimal delay values.

The circuit has a highly scalable structure, with the only source of variation being in the memory cell parameters to generate larger loop parameters. The local feedback wiring structure is meticulously designed so that each clock signal effectively removes the flux from the loop, generating an SFQ pulse. Subsequently, the loop undergoes a specific refresh time facilitated by the local feedback wiring. Thanks to the inherent speed of superconducting circuits, the recovery time required is primarily dictated by the timing of the CBU. This efficient design ensures rapid data refreshment and underscores the circuit's high-performance capabilities.

\section{Methodology and Design}
\begin{figure}[htp]
    \centering
    \begin{minipage}[t]{0.43\columnwidth}
        \centering
        \includegraphics[width=\linewidth]{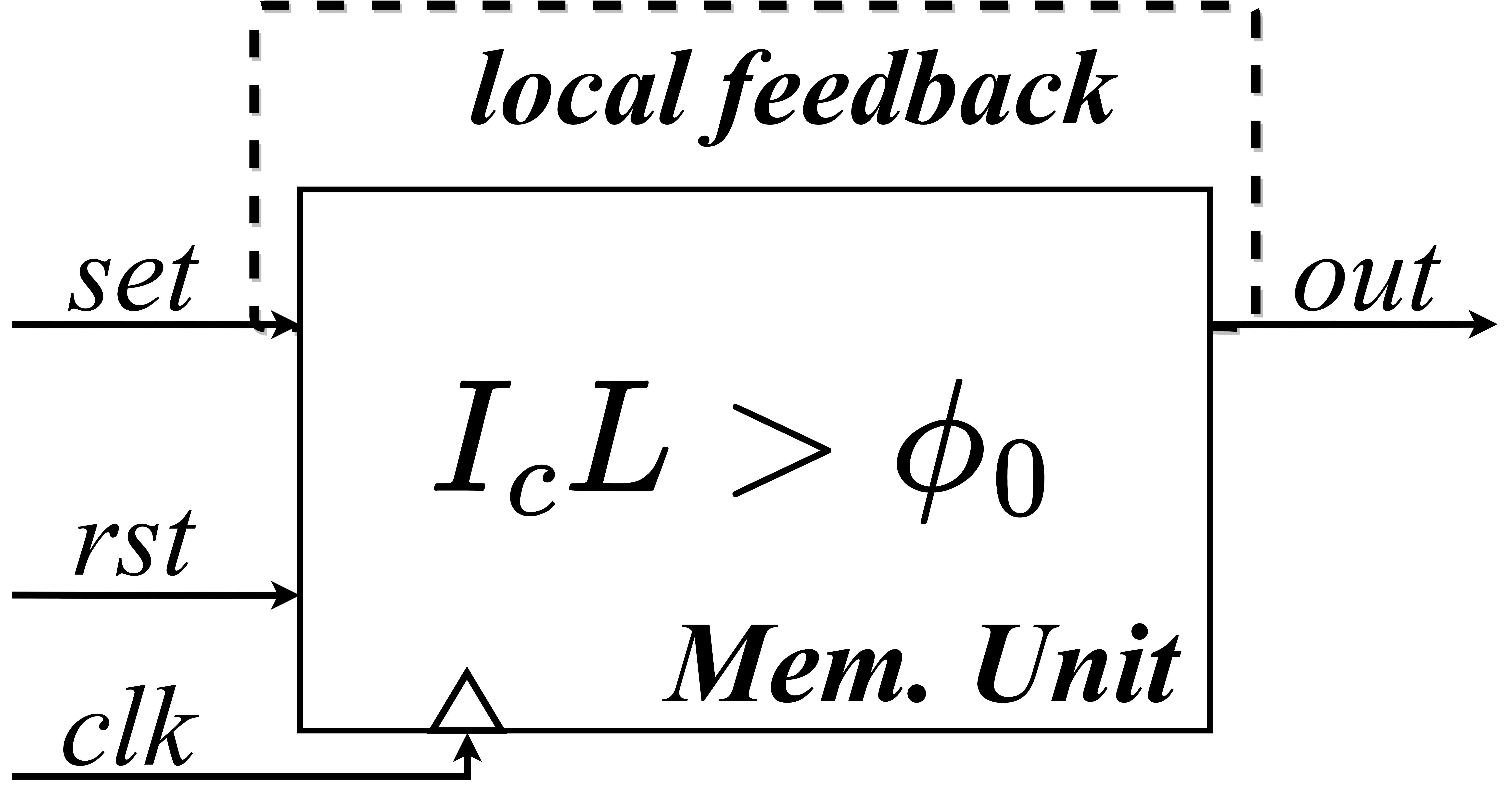}
        \captionsetup{font=small}
        \caption{NDRO block diagram}
        \label{fig:ndro}
    \end{minipage}%
    \hfill
    \begin{minipage}[t]{0.55\columnwidth}
        \centering
        \includegraphics[width=\linewidth]{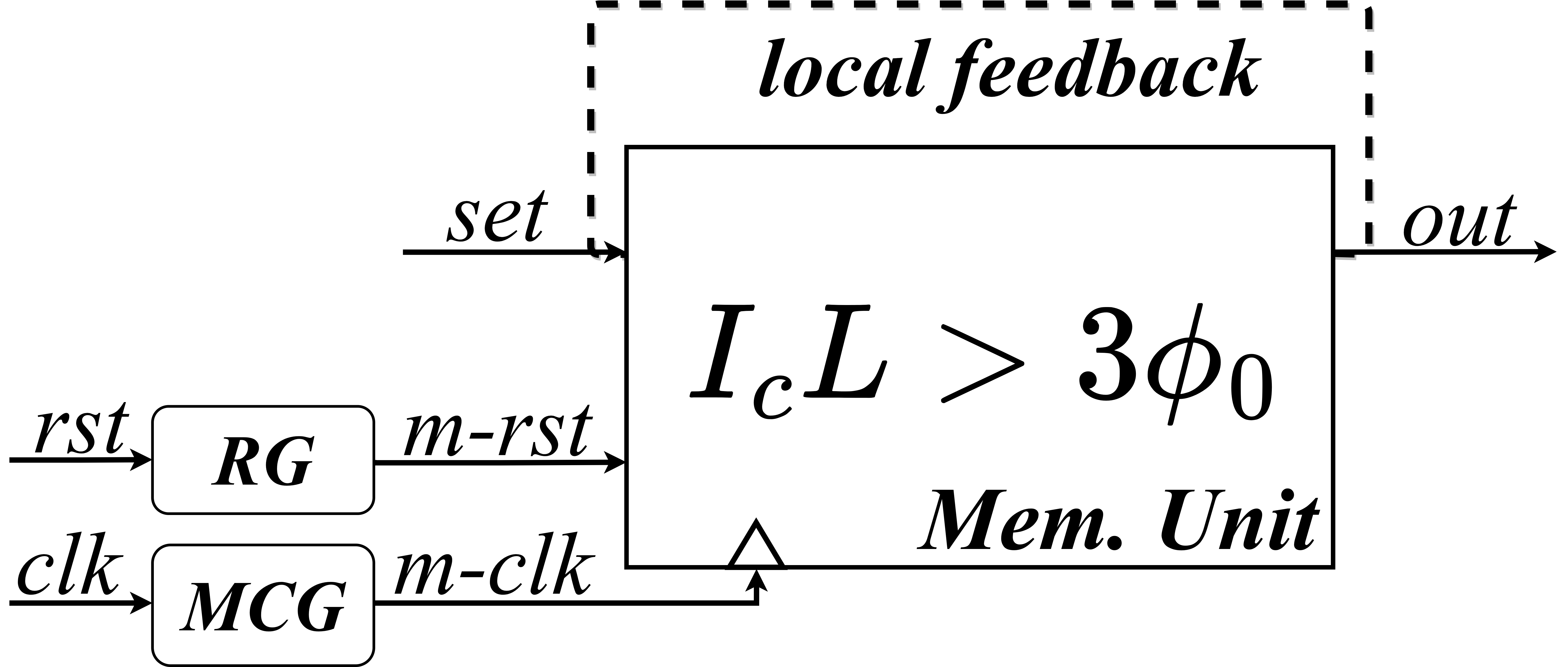}
        \captionsetup{font=small}
        \caption{M-NDRO block diagram}
        \label{fig:m-ndro}
    \end{minipage}
    \hfill
\end{figure}

\begin{figure}[htp]
    \begin{subfigure}{\columnwidth}
        \centering
        \includegraphics[width=0.5\linewidth]{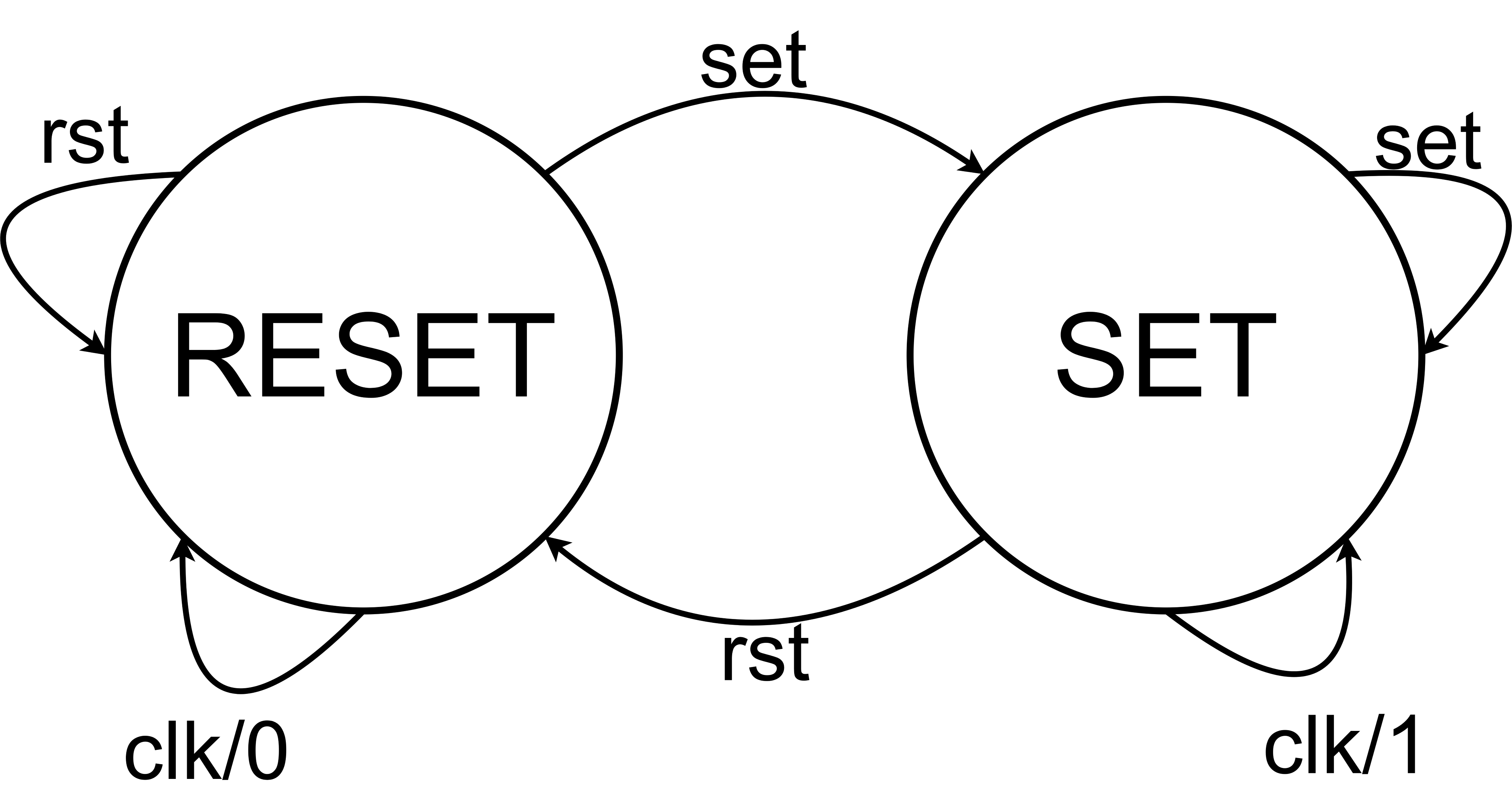}
        \caption{The state machine of NDRO is demonstrated. If the machine is at the SET state, the clock signal will generate a pulse at the output. In the RESET state, the clock will not generate any output.}
        \label{fig:ndro_sm}
    \end{subfigure}
    
    \begin{subfigure}{\columnwidth}
        \centering
        \includegraphics[width=0.9\linewidth]{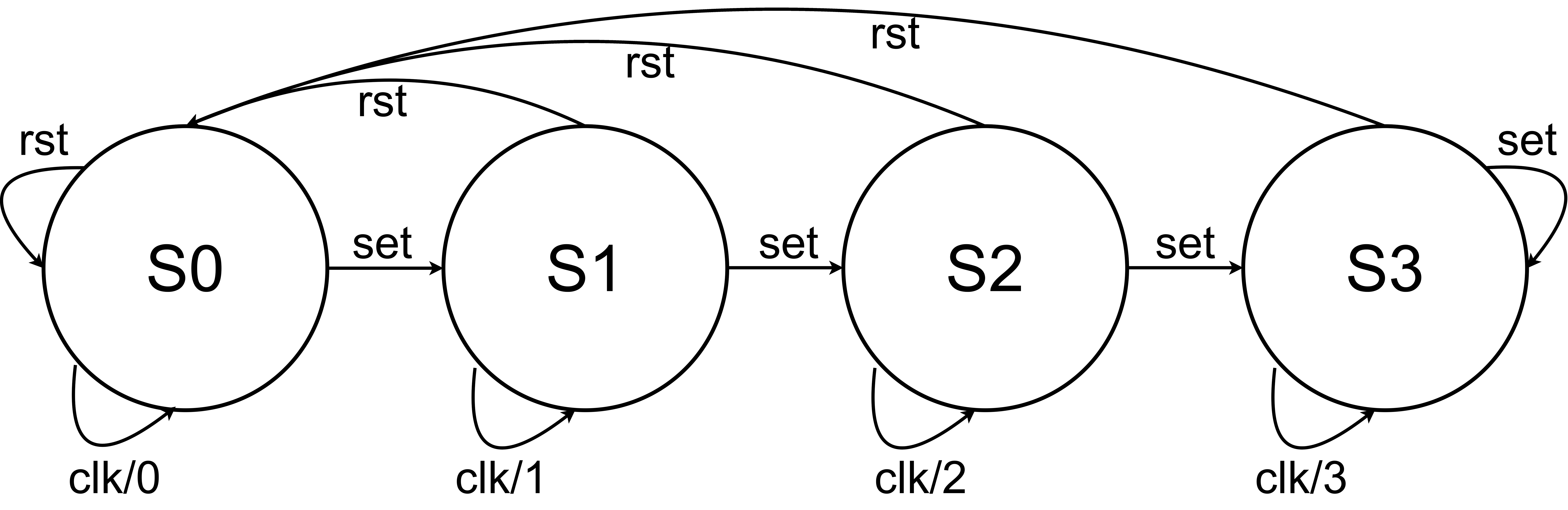}
        \caption{The state machine of M-NDRO with reset configuration. Each set signal increments the state until it reaches the final state and stays there. With the reset signal, the state goes back to zero. The clock signal generates one, multiple, or no pulses at the output based on the machine's state.}
        \label{fig:mndro_sm_rst}
    \end{subfigure}
    \begin{subfigure}{\columnwidth}
        \centering
        \includegraphics[width=0.9\linewidth]{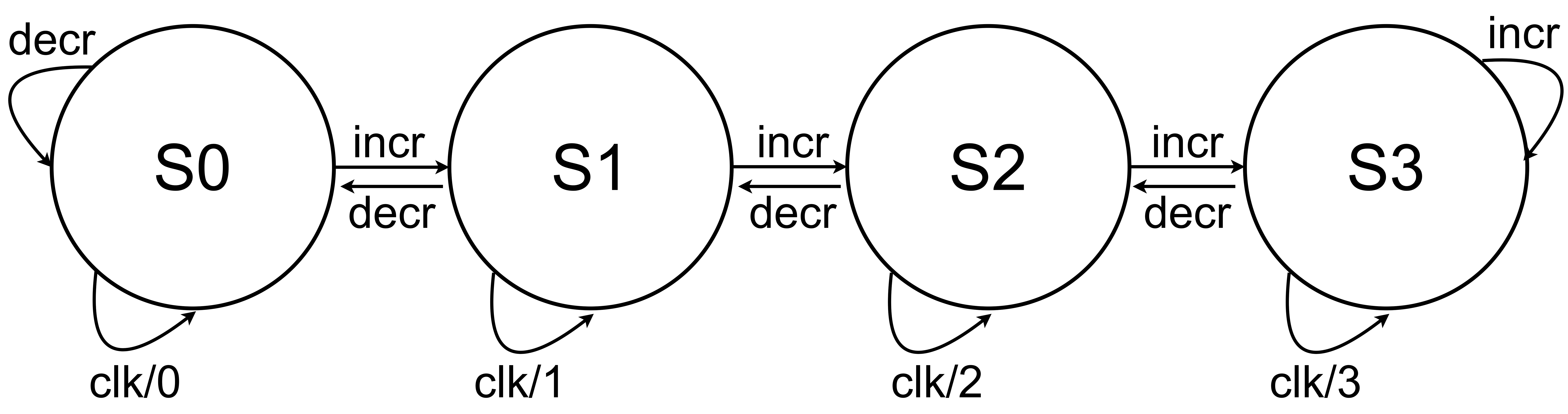}
        \caption{The state machine of M-NDRO with decrement configuration that acts as an up/down counter.}
        \label{fig:mndro_sm_dec}
    \end{subfigure}
    \caption{NDRO, M-NDRO State Machines}
\end{figure}

There are two main sub-circuits for the proposed NDRO design. The first is the memory unit for data storage, which has input pins of set, reset, and clock. The second part is local feedback wiring for reloading the data back into the memory unit.
The NDRO and M-NDRO block designs are shown in Fig. \ref{fig:ndro} and Fig. \ref{fig:m-ndro}, respectively.
In both configurations, the memory unit comprises a storage loop that stores 1-bit or 2-bit data. Although the structural framework of the memory unit remains consistent between these designs, differences arise in the parameters of flux storage loop values. The local feedback mechanism is generated through the implementation of wiring cells in both instances. Conversely, in the M-NDRO design, the requirement arises for multiple clock signal generation and reset generator units to facilitate the generation of the desired output. The absence of a reset generator unit results in the reset pin acting as a decrement function. Therefore, set and reset pins are increment and decrement pins. This structure can be used as an up/down counter, as a synapse circuit supporting 2-bit weight values in neural network implementations. The state machine demonstrating the NDRO design is illustrated in Fig.\ref{fig:ndro_sm}. In contrast, the state machines corresponding to the M-NDRO configuration, with and without the presence of a reset generator unit, are depicted in Fig.~\ref{fig:mndro_sm_rst} and Fig.~\ref{fig:mndro_sm_dec}.
The NDRO configuration embodies a binary state machine characterized by two distinct states labeled "RESET" and "SET." These states are subject to transition inputs initiated by the RST and SET pins. Activation of the RST pin leads to a transition to the "RESET" state, while the SET pin induces a change to the "SET" state.
In the "SET" state, applying the clock signal results in generating a single SFQ (Single Flux Quantum) pulse. Conversely, in the "RESET" state, the clock signal produces no pulse output, aligning with the expected behavior of a non-destructive memory unit.

The M-NDRO unit, when equipped with the RG Unit, features a state machine encompassing states denoted as S0 through S3, as illustrated in Figure \ref{fig:mndro_sm_rst}. The SET input is responsible for incrementally advancing the state, while the RST input universally resets the state back to S0. Within each state, we are applying a clock pulse to generate a variable number of SFQ pulses: 0 in S0, 1 in S1, 2 in S2, and 3 in S3. This configuration enables the unit to have multifunctional capabilities, contingent upon the inputs received from SET, RST, and the clock signal.

A four-state state machine of the M-NDRO unit without the RG Unit, covering the states labeled S0 to S3, is demonstrated in Fig.~\ref{fig:mndro_sm_dec}. The operation of this unit revolves around two primary input pins: the SET pin and the RST pin. The SET pin functions as an increment function, and its effect on the unit's operation is state-dependent. Unless the machine is in the final state, when the SET input is applied,  it adds another $\Phi_0$ flux to the loop and increases the state. With each state transition, the output pulse count increments, illustrating the sequential nature of this state machine. The RST pin acts as a decrement function within this configuration. With each RST input, the state drops by one until it reaches S0, where it remains there.

\subsection{Destructive Memory Unit Design}
The D-Flip Flop (DFF) is a conventional Destructive Readout (DRO) unit that stores a single bit of information. Its fundamental storage structure comprises three Josephson Junctions (JJ) and three ports: data in, clock, and data out. The output junction of the DFF switches with each clock signal, enabling its widespread application as a memory unit. Moreover, the DFF possesses a flexible architecture that allows for easy modification of the storage capacity by adjusting the critical current ($I_c$) and loop inductance ($\mathrm{L_{loop}}$) parameters. Unlike the DRO, the Non-Destructive Readout (NDRO) needs a reset pin to clear the stored value within the memory loop. 
\begin{figure}[htb]
  \includegraphics[width=1\linewidth]{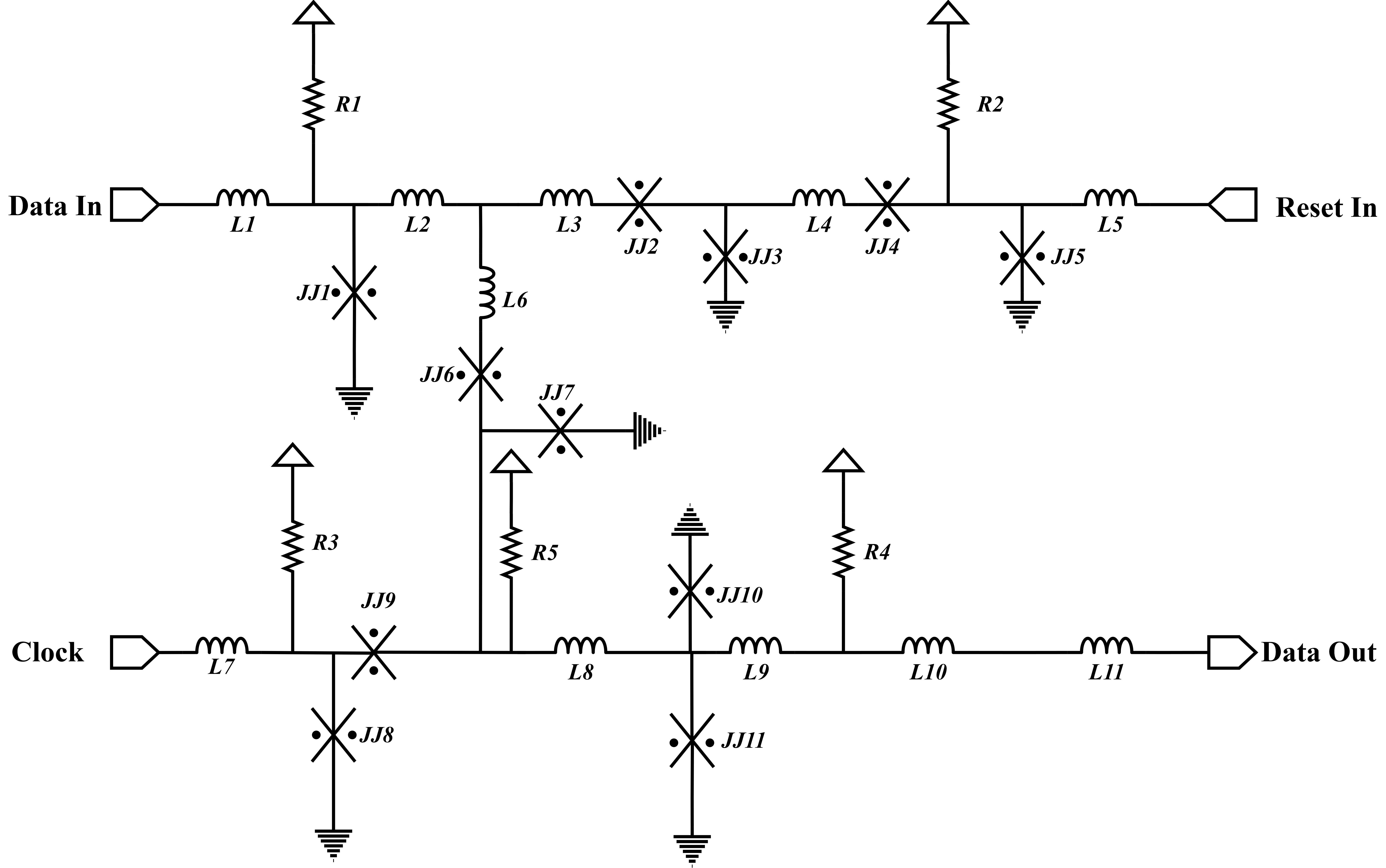}
  \caption{The schematic of set/reset flip-flop Memory Unit.}
  \label{fig:ndroSch}
\end{figure}
The design of a set/reset flip-flop, also known as RDFF structure, is essential for an NDRO cell design. The RDFF structure is similar to the method proposed in \cite{rdff_sch}. In the given design demonstrated in Fig.\ref{fig:ndroSch}, the RDFF unit storage loop is considered JJ1-L2-L6-JJ6-JJ7. The JJs and storage loop inductance values should be determined so that the loop capacity is higher than one flux quanta. The set pin stores the value; however, the reset pin breaks the loop and deflux it, sending the stored value to the ground. In the given design, the clock signal destroys the storage loop. However, we want our unit to provide the set value until the reset signal comes. Therefore, we need local feedback wiring to load the memory in a refreshed state.

\subsection{M-NDRO Memory Unit Design}
The proposed M-NDRO design has a storage loop and local feedback, just like the NDRO design. However, the memory unit for M-NDRO stores up to 3 SFQ pulses to represent 2 bits. Since the storage unit can store more than one fluxon, the clock should provide more than one output pulse. Also, reset should destroy all storage values, not just one pulse. Therefore, in the M-NDRO design, we have two additional modules called multiple clock generation (MCG) and reset generation (RG) units.

To address the limitations of storing only one pulse in a traditional RDFF, modifications were made to the circuit parameters to satisfy the criterion of $I_c\times L > 3\phi_0$, where $\phi_0$ denotes a quantum of electromagnetic flux, $I_c$ is the critical current of the loop JJs and L is the loop inductance. Accommodating a higher inductance value is imperative for storing more Single Flux Quantum (SFQ) pulses effectively. The schematic depiction of the NDRO and M-NDRO, based on the RDFF structure, is presented in Fig. \ref{fig:ndroSch}. The schematic for both designs is the same, but the circuit parameters are different. The main difference between the circuits is to store a higher number of SFQ pulses, the storage inductance values and critical current of the junctions should also increase. JJ1-L2-L6-JJ6-JJ7 parameters are crucial to determine the storage value. The parameters of M-NDRO are demonstrated in Table~\ref{table:MNDRO}.
\begin{table}[ht]
\centering
\caption{Parameter values of the NDRO and M-NDRO cells.} 
\label{table:MNDRO}
\begin{tabular}{|c|c|c|c|c|c|}
\hline
     Parameter&NDRO&M-NDRO&Parameter&NDRO&M-NDRO\\
     \hline
     L1 & 2.09pH & 1.89pH&
     L2 & 2.28pH& 7.36pH\\
     L3 & 2.84pH&2.14pH &
     L4 & 1.85pH&2.33pH\\
     L5 & 2.55pH & 4.10pH&
     L6 & 1.9pH & 3.91pH\\
     L7 & 1.94pH & 1.73pH&
     L8 & 1.79pH & 0.66pH\\
     L9 & 1.90pH & 1.71pH&
     L10 & 0.58pH & 0.45pH\\
     L11 & 2.20pH & 1.84pH&
     J1 & $290\mu A$ & $382\mu A$\\
     J2 & $208\mu A$ & $311\mu A$&
     J3 & $248\mu A$ & $365\mu A$\\
     J4 & $361\mu A$ & $341\mu A$&
     J5 &  $245\mu A$ & $274\mu A$\\
     J6 &  $197\mu A$ & $158\mu A$&
     J7 &  $249\mu A$ & $256\mu A$\\
     J8 &  $247\mu A$ & $230\mu A$&
     J9  & $261\mu A$ & $263\mu A$ \\
     J10 &  $381\mu A$ & $372\mu A$&
     J11  & $209\mu A$ & $270\mu A$ \\
     R1 &  10.89 $\Omega$ & 5.42 $\Omega$&
     R2  & 10.17 $\Omega$ & 7.59 $\Omega$ \\
     R3 &  11.47 $\Omega$ & 10.34 $\Omega$& 
     R4  & 7.99 $\Omega$ & 8.67 $\Omega$ \\
     R5 &  10.33 $\Omega$ & 9.17 $\Omega$& & & \\
\hline
\end{tabular}
\end{table}
\subsection{Local Feedback Wiring}
\begin{figure}[htb]
\centering
  \includegraphics[width=0.5\linewidth]{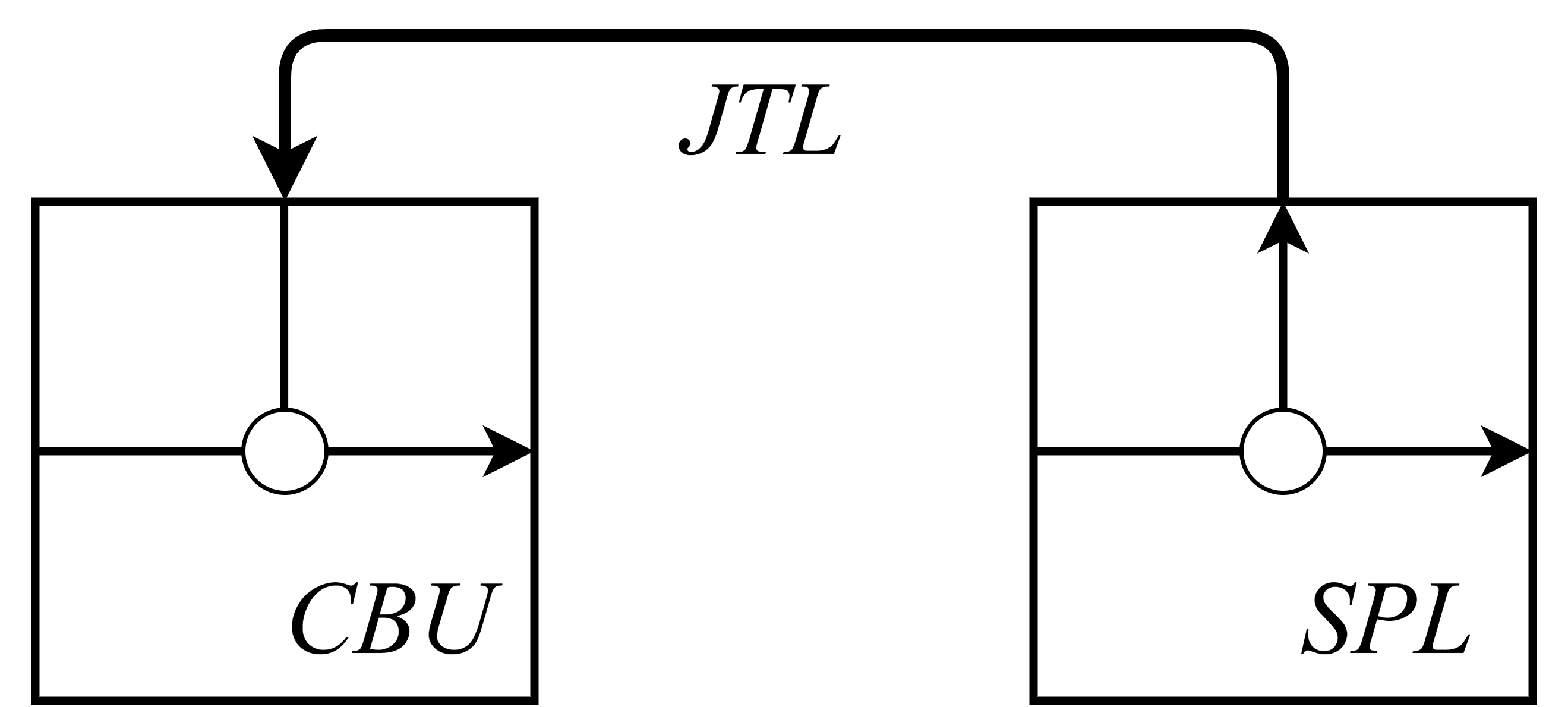}
  \caption{Local Feedback Wiring Block Diagram.}
  \label{fig:localWiring}
\end{figure}

In RSFQ circuits, there are two types of wiring cells: passive and active. Active wiring cells consist of Josephson Junctions for transmitting the data. The JTL, SPL, and CBU are considered active wiring cells. Since RSFQ has a unit fan-in/fan-out capability, we must split and merge the signals for using them in different circuits. In the proposed NDRO design, we want to reload our output to the input to provide non-destructive behavior. The SPL functions to split the output of the memory unit to the output of the NDRO and the input of the memory unit. The CBU merges the input and feedback signals and supplies them to the memory unit. 
Also, the timing values and the JJ counts of the wiring cells are given in Table \ref{tab:delayTable}.
 \begin{table}[ht]
\centering
\caption{Wiring Cells Specifications}
\label{tab:delayTable}
\begin{tabular}{|c|c|c|c|}
\hline
 & JTL  & SPL & CBU \\
\hline
In-to-Q Delay (ps) & 3   & 2.5   & 5   \\
\hline
JJ Count & 2   & 3   &  7   \\
\hline
\end{tabular}
\end{table}
Delay values for the wiring cells are fast enough to provide a fast reload to the NDRO memory unit. Each local wiring needs one of each wiring cell. The reloading time takes 10.5ps. The basic block diagram of the local feedback wiring circuit is given in Fig. \ref{fig:localWiring}.
Upon the arrival of the clock signal, the circuit generates an output if a stored flux is present in the loop. Following this generation, the output is split using an SPL cell, which features one input and two identical outputs. One output is directed to the output port of the NDRO, while the other is routed to the JTL for reloading the stored value back into the memory unit. The JTL's output serves as one input to a CBU, with two inputs and one output. One input is dedicated to reloading the stored value, and the other input is employed as the set input for the NDRO cell. Depending on the delay values outlined in the table, the pulse is reloaded with a 10.5ps delay following the defluxing of the loop triggered by the clock input.
The local feedback wiring structure introduces 12 Josephson Junctions (JJs) to the circuit. The same local feedback circuit is employed in the M-NDRO cell for various storage capacities. Consequently, the local feedback overhead remains constant in the M-NDRO circuit and does not increase with the expansion of storage capabilities.

\subsection{Multiple Clock Generator Circuit Design}
In conventional synchronous circuits, a clock signal typically triggers the output junction, generating a single SFQ pulse at the output port. However, in an M-NDRO circuit, the number of pulses in the memory unit can vary depending on its state. The output junction must be triggered multiple times to obtain the desired output in such cases.
To maintain compatibility and consistency with other circuits that rely on a single SFQ pulse for the clock input, we employ a Multiple Clock Generator (MCG) circuit. In the worst-case scenario, this circuit can generate up to three SFQ pulses from a single clock pulse, ensuring that the required number of pulses is available to drive the M-NDRO unit effectively.

\begin{figure}[htb]
\centering
  \includegraphics[width=1\linewidth]{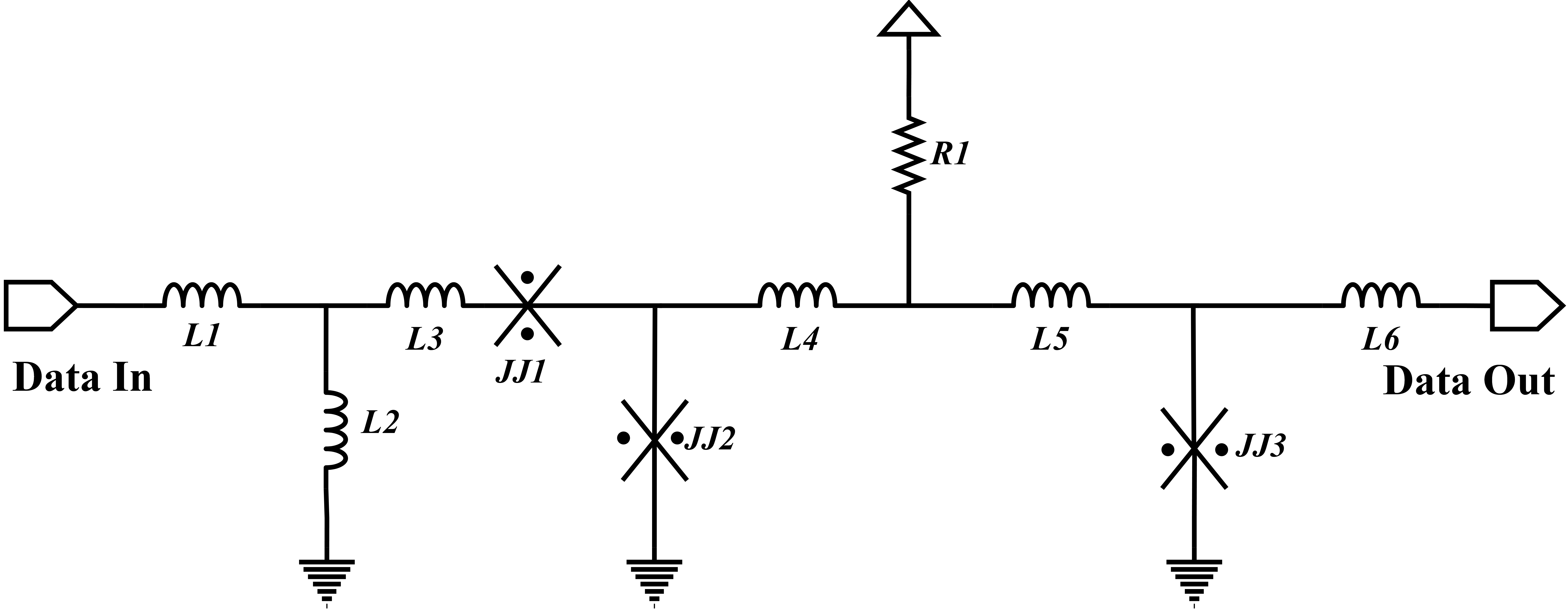}
  \caption{The schematic of MCG circuit. The circuit generates multiple SFQ outputs from a single input pulse.}
  \footnotesize(L1=0.6pH, L2=7pH, L3=2.28pH, L4=0.43pH, L5=2.86pH, L6=4.05pH, JJ1=170$\mu$A, JJ2=150$\mu$A, JJ3=230$\mu$A, R1=7.2$\Omega$)
  \label{fig:dcSfqSch}
\end{figure}

\begin{figure}[htb]
\centering
  \includegraphics[width=1\linewidth]{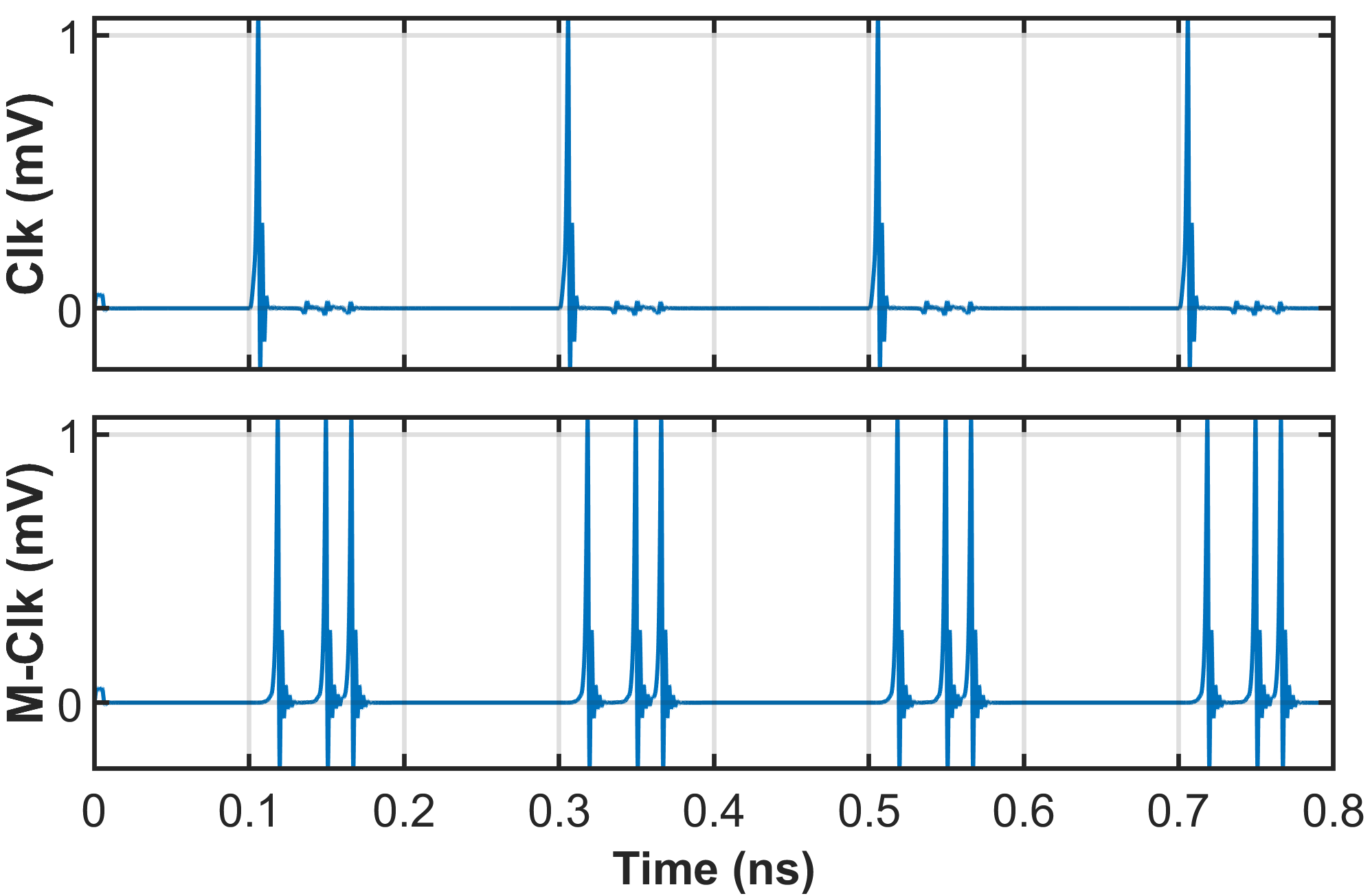}
  \caption{The simulation result of the MCG circuit.}
  \label{fig:dcSfqSim}
\end{figure}

As illustrated in Figure \ref{fig:dcSfqSch}, the MCG circuit utilizes a DC-SFQ converter structure that essentially operates as a threshold gate. By carefully adjusting the critical currents of the junctions (JJ1 and JJ2), the threshold value of the circuit can be modified, enabling the generation of up to three SFQ pulses from a single input pulse. This circuit includes a single clock-in port and one multiple clock-out port.
The circuit configuration involves three JJs specifically designed to produce multiple SFQ pulses from a single input pulse efficiently, thus expanding the circuit's functionality for applications where such pulse multiplication is necessary.
In the simulation depicted in Figure \ref{fig:dcSfqSim}, when a single clock-in pulse is delivered from the clock port of the M-NDRO circuit, the MCG circuit effectively produces three pulses. This identical circuit configuration is also employed in the reset generator unit for a specific reason.
The decision to use this configuration in the reset generator unit is driven by the necessity to replicate the reset signal like the clock signal. In SFQ-based systems, each stored SFQ pulse requires its reset pulse to ensure proper clearing. Therefore, to accommodate the worst-case scenario, where the reset case involves the storage of 3 SFQ pulses, generating three reset signals from a single reset input is imperative. This approach guarantees a dependable and robust reset operation.
The SPL3 and CBU3 circuits are also suitable for the multiple clock and reset generation. They can generate the 3 SFQ pulses from 1 SFQ pulse but are not as compact as the proposed circuit.
\section{Simulation Results}
\subsection{NDRO Simulation Results}
\begin{figure}[htp!]
\centering
  \includegraphics[width=1\linewidth]{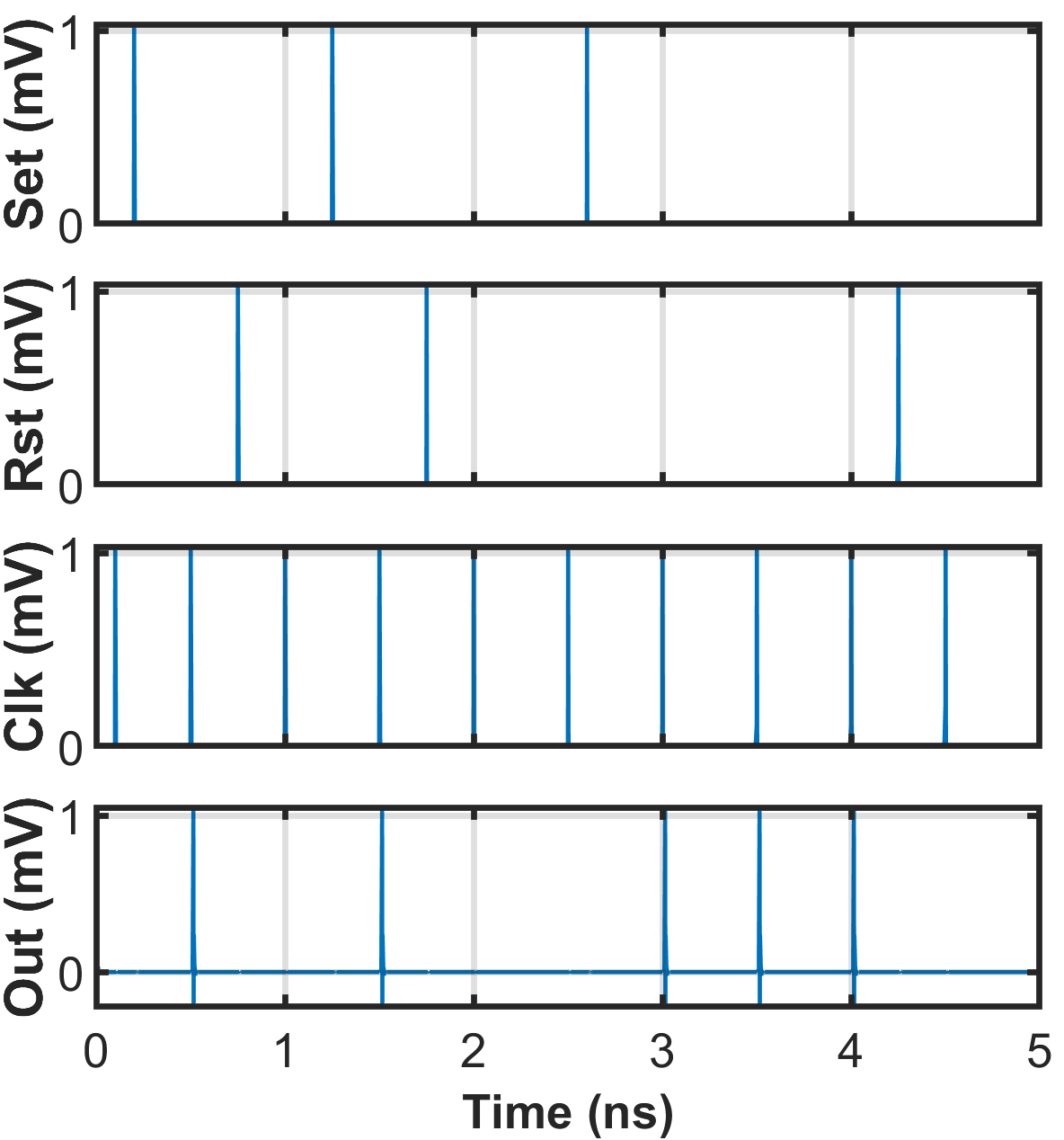}
  \caption{The simulation result of the NDRO circuit. The simulations are done by using the JSIM software.}
  \label{fig:ndroSim}
\end{figure}

A detailed evaluation of the NDRO circuit's operational performance involved testing it under various input scenarios. As illustrated in Figure \ref{fig:ndroSim}, the NDRO circuit exhibited the expected behavior and operated with precision.
The circuit persisted the predetermined set value until the reset signal was triggered, aligning with its non-destructive memory function. Furthermore, the circuit generated an SFQ pulse with each clock signal based on the information stored in the set configuration.
For example, in the initial time window depicted in the figure, the NDRO circuit was preloaded with a predefined set value. Upon the arrival of the clock signal, the output junction promptly fired, demonstrating successful operation. When the reset signal was activated in the subsequent time window, a destructive event occurred, resulting in the absence of an output pulse, which is the expected behavior of a reset operation.
However, in the following time window, just before the arrival of the reset signal, the circuit performed as anticipated, displaying the loaded input value with each occurrence of the clock signal. 
\subsection{M-NDRO Simulation Results}
\begin{figure}[htb]
\centering
  \includegraphics[width=1\linewidth]{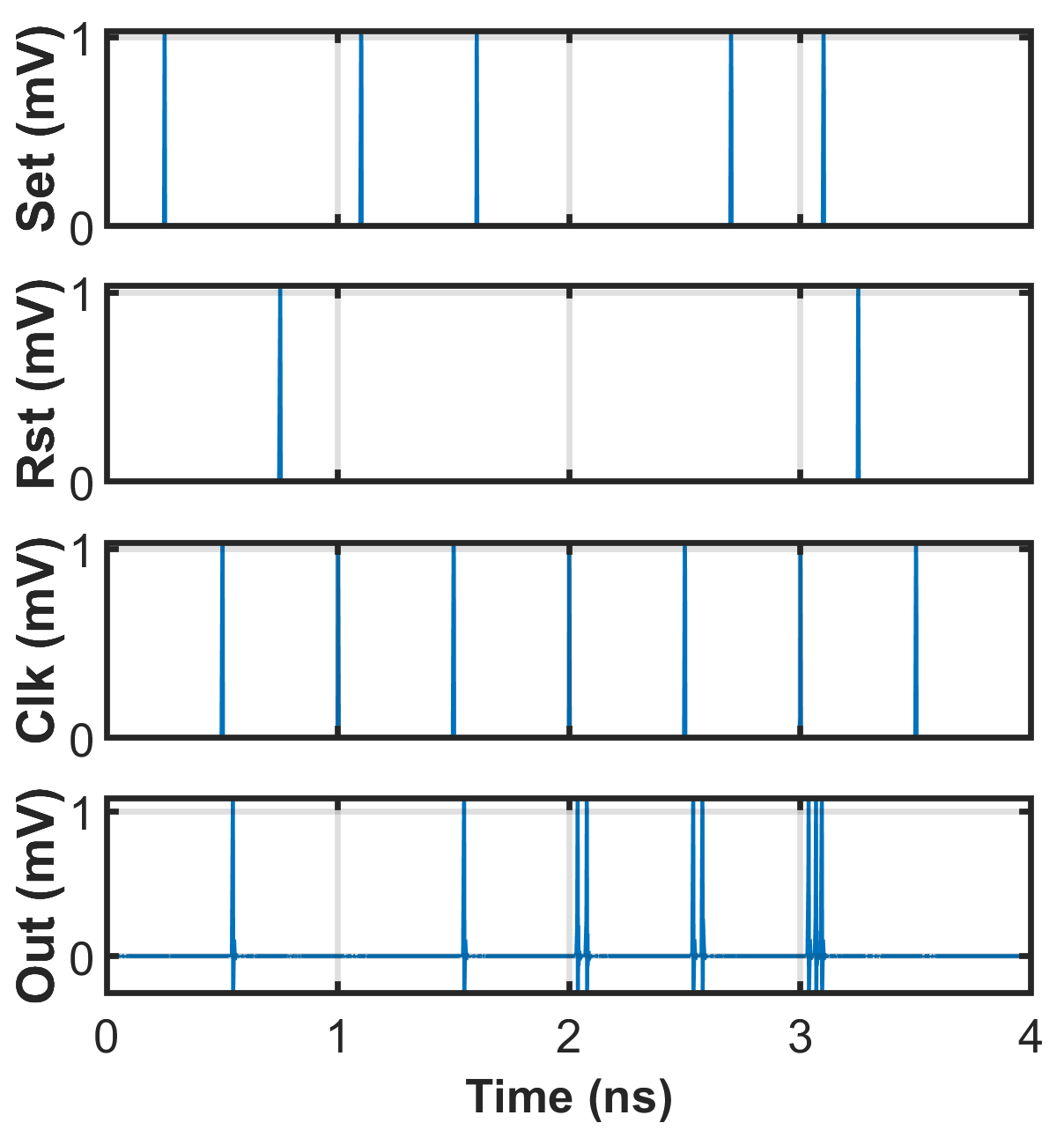}
  \caption{The simulation result of the M-NDRO circuit with reset configuration.}
  \label{fig:mndroSim}
\end{figure}

The M-NDRO circuit comprises a high flux storage unit with multiple clock and reset generators. The circuit's behavior has been accurately simulated and is visualized in Figure \ref{fig:mndroSim}.
The primary objective of the M-NDRO circuit is to retain up to 2-bit values within the storage loop and deliver them to the output port in sequence upon the arrival of the clock signal.

In the initial simulation case, a single SFQ pulse was stored in the M-NDRO circuit. Subsequently, the reset signal (RST) cleared the loop in the following time window, returning it to its initial state. In the subsequent time frame, two SFQ pulses were observed to be successfully stored within the M-NDRO circuit. At the subsequent clock cycle, another set signal was applied, and the M-NDRO accommodated three SFQ pulses, indicating its ability to store multiple bits of information. However, it's essential to note that no output is observed when the reset signal arrives, as illustrated in the following time frame. This absence of an output signifies that the storage unit had been completely cleared of its flux, aligning with the expected behavior.

The M-NDRO circuit's functionality, as showcased in the simulation, demonstrates its capability to accumulate and sequentially deliver multiple SFQ pulses with a clock signal while ensuring reliable readout and reset operations. The circuit achieves a maximum clock frequency of 10GHz, although it's important to note that the multiple clock generation overhead influences this frequency.

\begin{figure}[htb]
\centering
  \includegraphics[width=1\linewidth]{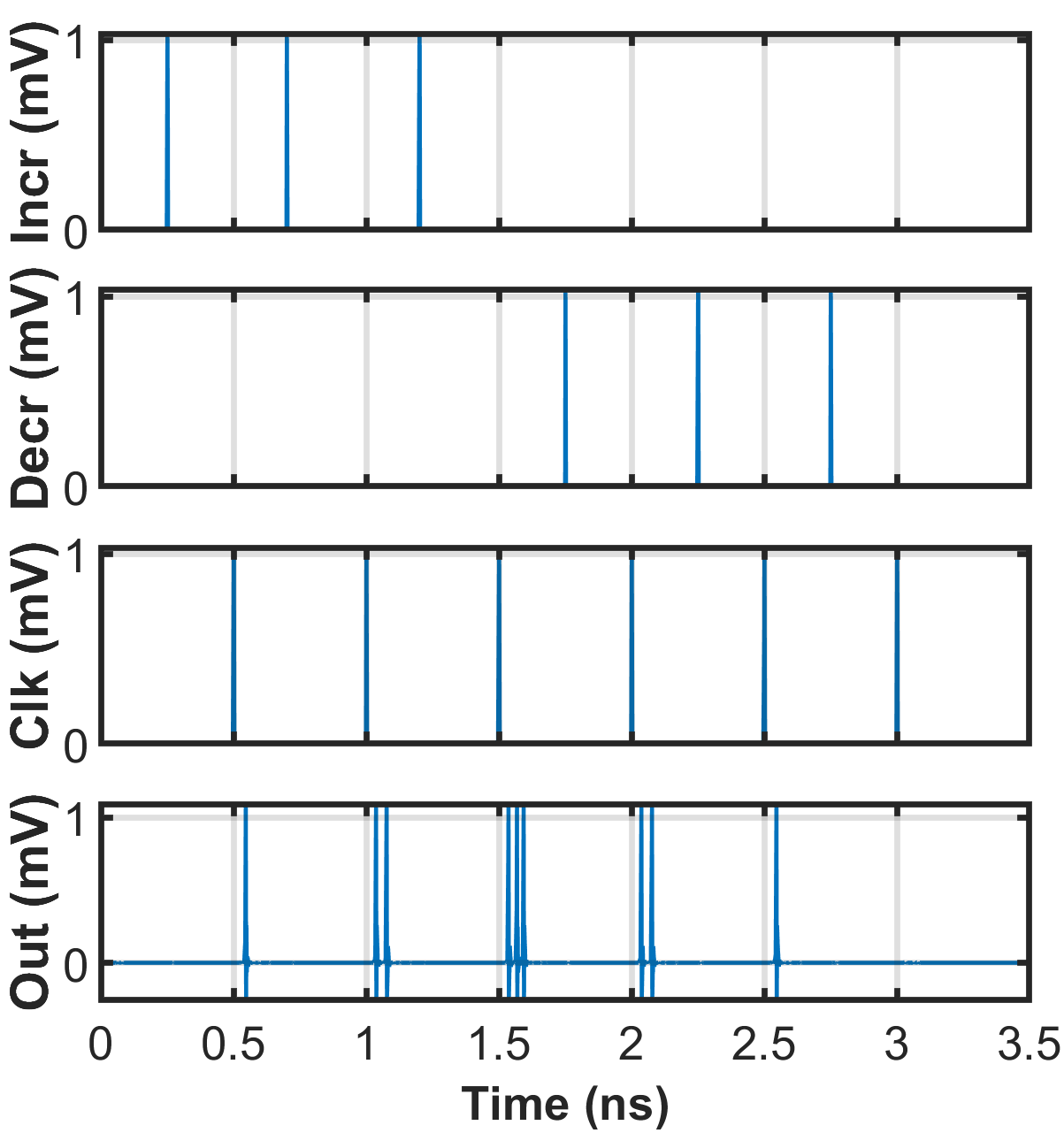}
  \caption{The simulation result of the M-NDRO circuit with increment/decrement properties.}
  \label{fig:mndroSimIncr}
\end{figure}

In Figure \ref{fig:mndroSimIncr}, we look into the increment and decrement features of the circuit design. In this configuration, the set pin effectively functions as an increment input, while the RST pin operates as a decrement input. Without the Reset Generator (RG), the RST input removes only one SFQ pulse at a time, resulting in a decrement operation. The simulation results depicted in Figure \ref{fig:mndroSimIncr}  illustrate the circuit's capability to incrementally count from 0 to 3 and then decrease from 3 back to 0. 
\section{Margin Analysis}
The margin results for both the NDRO and M-NDRO cells were obtained through extensive simulations using the qCS software. These simulations allowed us to determine critical parameters for the storage loops of both NDRO designs. 
The NDRO cell's margin range was calculated to be an impressive 64\%, indicating a high resilience against process variations. In contrast, for the M-NDRO cell, the margin range was measured at 20\%, which, while slightly lower than that of the NDRO, still represents a substantial margin and signifies its robustness in practical applications. These margin values underscore the reliability of both the NDRO and M-NDRO circuits.

After the analysis, critical parameters are found on the storage loop of NDRO designs. High margins on the design parameters represent the reliability against process variations. The margin range for the NDRO is measured as 64\%, while it is measured as 20\% for the M-NDRO. 

\section{Conclusion}
In this study, we introduced and presented the designs of two memory units: a 1-bit Non-Destructive Readout (NDRO) and a 2-bit Multi-NDRO (M-NDRO) cell, both featuring a local feedback wiring structure. Despite their structural similarities, these memory cells offer distinct flux storage capabilities achieved by varying their respective parameter values.
The proposed circuit demonstrates a scalable architecture that can adapt to different parameter sets, allowing for the generation of higher storage capacities. Additionally, the circuit can increment and decrement magnetic flux storage, rendering it useful as a counter.
Operating at a frequency of 10GHz and boasting a substantial 20\% parameter margin for the M-NDRO cell, the proposed design enables the development of denser and high-performance cryogenic memory devices. This work represents a significant step forward in memory unit design, with potential applications in many cryogenic computing systems.

\section{Acknowledgments}
This work is supported by a grant from the National Science Foundation (NSF) under the project Expedition: Discover (Design and Integration of Superconducting Computation for Ventures beyond Exascale Realization) grant number 2124453.
The authors want to thank Haolin Cong for his contributions.

\end{document}